\documentclass[a4paper,10pt, conference]{IEEEtran}
\usepackage[left=1.57cm,right=1.57cm,top=0.95cm,bottom=2.54cm]{geometry}
\IEEEoverridecommandlockouts
\usepackage{cite}
\usepackage{amsmath,amssymb,amsfonts}
\usepackage{algorithmic}
\usepackage{graphicx}
\usepackage{textcomp}
\usepackage[utf8]{inputenc}
\usepackage{enumerate}
\usepackage{algorithm}
\usepackage{algorithmic}
\usepackage{booktabs}
\usepackage{xcolor}

\usepackage[hyphenbreaks]{breakurl}

\def\BibTeX{{\rm B\kern-.05em{\sc i\kern-.025em b}\kern-.08em T\kern-.1667em\lower.7ex\hbox{E}\kern-.125emX}}

\begin{document}


\title{Urban Traffic Monitoring and Modeling System:
\linebreak An IoT Solution for Enhancing Road Safety}

\author{
	\IEEEauthorblockN{
	Rateb Jabbar\IEEEauthorrefmark{1}\IEEEauthorrefmark{2},
	Mohammed Shinoy\IEEEauthorrefmark{1}, 
	Mohamed Kharbeche\IEEEauthorrefmark{1}, 
	\\
	Khalifa Al-Khalifa\IEEEauthorrefmark{1},
	Moez Krichen\IEEEauthorrefmark{3},
	Kamel Barkaoui\IEEEauthorrefmark{2}
	}
	\\
	\IEEEauthorblockA{
	\textit{\IEEEauthorrefmark{1}Qatar Transportation and Traffic Safety Center, Qatar University, Qatar} \\
	{\{rateb.jabbar,m.shinoy,mkharbec,alkhalifa\}}@qu.edu.qa \\
	}
	
	   \\
	\IEEEauthorblockA{
	    \textit{\IEEEauthorrefmark{2}Cedric Lab, Computer Science Department,Conservatoire National des Arts et Meteirs, France } \\
	    {\{jabbar.rateb.auditeur,kamel.barkaoui\}}@cnam.fr
	}
	\\
	\IEEEauthorblockA{
	    \textit{\IEEEauthorrefmark{3}ReDCAD Laboratory, National School of Engineers of Sfax, University of Sfax, Tunisia }\\
		{moez.krichen}@redcad.org \\
    } 
}

\IEEEoverridecommandlockouts
\IEEEpubid{\makebox[\columnwidth]{978-1-7281-5184-7/19/\$31.00~\copyright2019 IEEE \hfill} \hspace{\columnsep}\makebox[\columnwidth]{ }}


\maketitle

\begin{abstract}

Qatar expects more than a million visitors during the 2022 World Cup, which will pose significant challenges. The high number of people will likely cause a rise in road traffic congestion, vehicle crashes, injuries and deaths. To tackle this problem, Naturalistic Driver Behavior can be utilised which will collect and analyze data to estimate the current Qatar traffic system, including traffic data infrastructure, safety planning, and engineering practices and standards. In this paper, an IoT-based solution to facilitate such a study in Qatar is proposed. Different data points from a driver are collected and recorded in an unobtrusive manner, such as trip data, GPS coordinates, compass heading, minimum, average, and maximum speed and his driving behavior, including driver’s drowsiness level. Analysis of these data points will help in prediction of crashes and road infrastructure improvements to reduce such events. It will also be used for drivers' risk assessment and to detect extreme road user behaviors. A framework that will help to visualize and manage this data is also proposed, along with a Deep Learning-based application that detects drowsy driving behavior that netted an 82\% accuracy.

\end{abstract}


\section{Introduction}
According to the Global status report~\cite{who2019} on road safety 2018 from the World Health Organization, Road crashes and related forms of crashes is the eighth leading cause of death globally. According to the crash data published in Qatar for 2017~\cite{ministry2018}, the total number of road crashes in Qatar is around 240,333 ~\cite{ministry2018} and negligence and reckless driving come as one of the first leading causes. It is worth to mention that 97.5\% of these crashes are minor.

Qatar expects more than a million visitors during the 2022 World Cup, which will pose significant challenges. The high number of people will likely cause a rise in road traffic congestion and safety concerns. To tackle this problem, a system that incorporates Naturalistic Driving Behavior (NDB) can be utilized to collect and analyze data that help to understand the current Qatar traffic system and provide suggestions for better infrastructure, safety planning, and engineering interventions. 

In order to understand driver behavior, NDB Study is used to investigate  the  causes  of  crashes  and  the  typical  daily  driving behavior. There is no accepted definition of Naturalistic Driving Study. However, as defined by \cite{udrive}, it is in general considered as an unobtrusive observation method for investigating daily driving behavior of drivers in a non-experimental and natural setting (Schagen et al., 2012~\cite{VanSchagen2012}; Dingus et al., 2014~\cite{Dingus2015}; Ingrid et al., 2009~\cite{Ingrid2010};). In this kind of study, drivers are not given instructions on how, where, and when to drive their vehicles. Accordingly, researchers can observe the natural interaction of drivers with traffic, roadway, and other vehicles during daily driving activities. Due to the improvements in storage capacities and data collection technology, researchers are able to carry out such studies on a large scale. In these studies, the standard Data Acquisition System (DAS) generally consists of video cameras, forward radars, lane tracking system, vehicle network, data storage system, eye-tracking system, Geographic Positioning System (GPS), vehicle network information, and accelerometers. To perform a naturalistic driving study, it is necessary to have a significant amount of data. As an example, in two years, SHRP2~\cite{campbell2012shrp} NDS obtained 2 petabytes (2 million gigabytes) data from over 3,000 drivers.

Furthermore, Machine-to-Machine (M2M)~\cite{Nahrstedt2016} is defined as the  communication between mobile devices, actuators, smart sensors, embedded processors, and computers without or with limited human intervention. M2M is applied in areas such as city automation, safety and security, transportation management, e-health, smart grid, and smart power. These devices are interconnected and known as the Internet of Things (IoT). The IoT enables intelligent monitoring, reporting, and control of different areas of our daily lives. IoT implementations are on the rise due to the advancements in internet speed with higher bandwidth and smaller telecommunication equipment.

\begin{figure*}[!t]
	\centering
	\includegraphics[width=.8\textwidth]{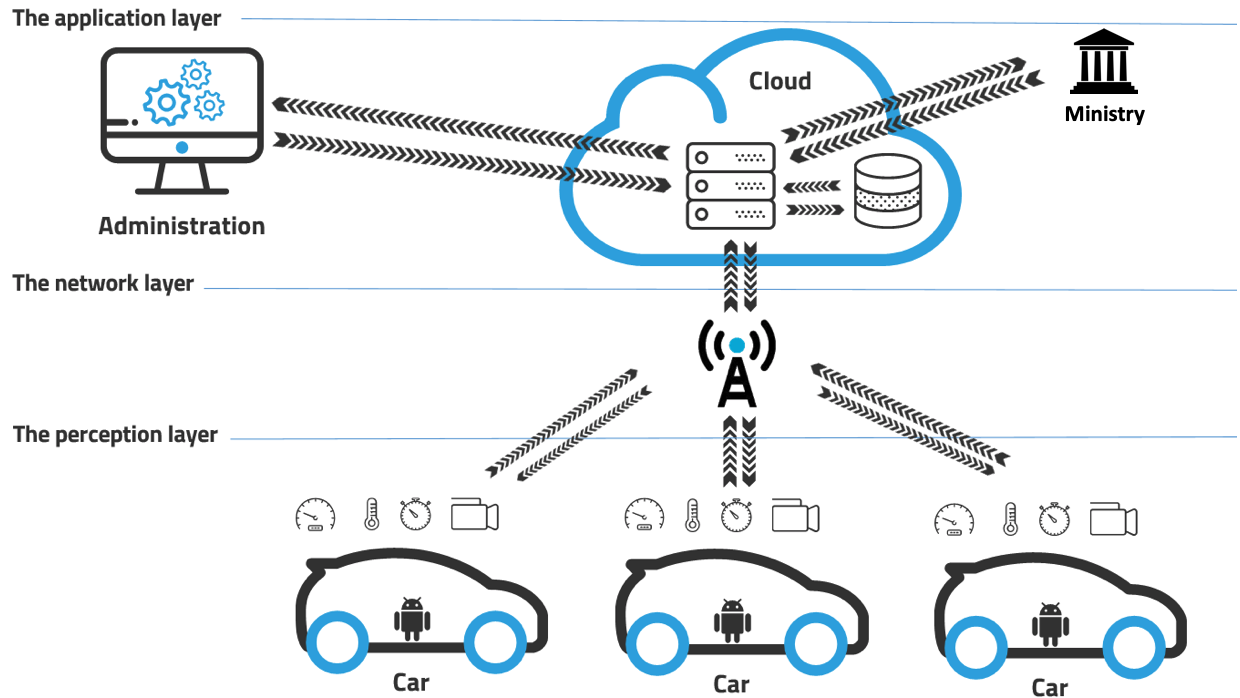}
	\caption{Architecture of the proposed Internet of things solution.}\label{architecture}
\end{figure*}

In this context, our proposed solution includes a system that captures driver behavior using mobile phone technologies. The system is able to capture the driver's GPS data and analyze facial image data to predict drowsiness using deep learning techniques such as CNN  to classify drowsy driving. Furthermore, this solution also enables the collection of data about every vehicle trip which facilitate the NDB studies.

The rest of this paper consists of five sections. First, an overview of the developed system is described in section II. Following that, the experimental results and analysis are discussed in section III. Next, in section IV, summary and main conclusions are presented. Finally, in Section V, Future work is presented.

\section{Methodology}

The architecture of our solution consists of three layers as outlined below: perception, network, and application layers:
\begin{enumerate}[i.]
	\item The perception layer consists of multiple sensors that are designed to sense and collect information about the surroundings, including physical parameters and identification of nearby smart objects.
	\item The network layer is in charge of connecting sensors to other servers, network devices, and smart things, as well as transmission and processing of sensor data.
	\item The application layer delivers application-specific services to the user. The architecture of the proposed solution is shown in figure .\ref{architecture}.
\end{enumerate}

\subsection{The perception layer}
In the perception layer, an Android application performs two functions: Firstly, a driver drowsiness detection which analyses the facial data and alerts the user regarding drowsy driving behavior. The second function is to collect data about the trip of the car

\subsubsection{Driver drowsiness detection}

The goal of the driver drowsiness detection system is to prevent crashes caused by drowsiness. This is an important step that will enhance the existing Advanced Driver Assistance Systems (ADAS) in a car. The role of ADAS is to improve safety and ensure a satisfactory driving experience for the driver.

In recent years, there has been accelerated advancement in machine learning, particularly in the field of deep learning. This advancement along with the boom in the usage of embedded smart devices has resulted in increased data collection and connectivity. This has contributed to building efficient solutions to improve the current driver drowsiness detection system.

Our work aims to develop a drowsiness detection system with improved efficiency based on CNN as a classifier on Android platforms.

\paragraph{Literature review}
Deep learning has become a widely used method for resolving challenging classification problems where conventional machine learning methods have lower accuracy. CNN's are used for tasks related to machine vision such as image classification, image segmentation, object detection ~\cite{He2015} \cite{Long}, etc. This technology has also been used by researchers to detect drowsy driving behavior too. 

Vijayan et al. \cite{vijayan2019} used 3 different CNN models to form a feature fused architecture to detect drowsiness which resulted in a 78\% accuracy. Dwivedi et al. \cite{Dwivedi2014} used shallow CNNs for detecting drowsy drivers with a 78\% accuracy rate. Liu, Weihuang, et al. \cite{liu2019} used a two-stream network along with multi-facial features to detect drowsiness on the NTHU-DDD dataset. The two-stream network was able to combine static and dynamic image information, this implementation even accounts for improving the lighting conditions of the images by using gamma correction on the images. 

Miguel et al. \cite{Garcia-Garcia2018} have implemented a low-cost drowsy detection system and it was able to achieve an accuracy of 72\%. This implementation shows promises as they were able to develop the system for an Android application. In our related work \cite{Jabbar2018}, we were able to attain an 81\% accuracy on average using an MLP model with facial landmarks while keeping the size of the model at only 100KB.

In this paper, our contribution is a CNN algorithm for drowsiness detection and a framework that can work with the mobile application to collect this data for naturalistic driving behaviour studies.

\begin{figure}[!t]
	\centering
	\includegraphics[width=.4\textwidth]{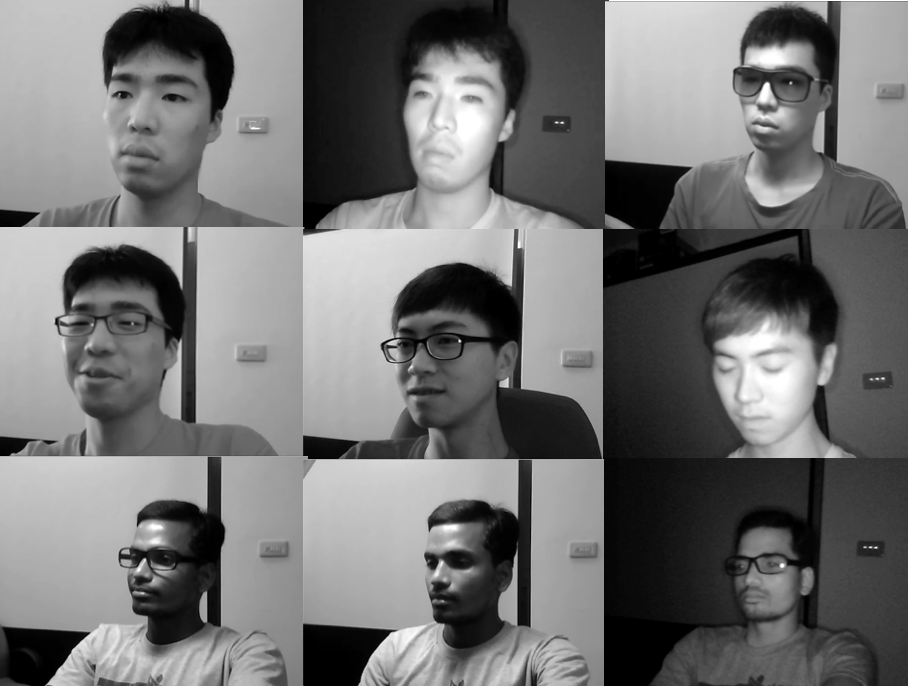}
	\caption{Sample of the pictures from the selected dataset}
	\label{fig:dataset_sample}
\end{figure}

\paragraph{Dataset and Preprocessing}
In this study, the National Tsing Hua University Driver Drowsiness Detection dataset is used\cite{Weng2017}. From this dataset, 22 participants of varied ethnic diversity are selected, a sample of which is given in figure \ref{fig:dataset_sample}. Out of this, 18 people were used in training data and 4 of them were used in the testing. All subjects were recorded in different simulated driving scenarios under night and day-lighting conditions. The participants enacted different driving behaviors such as dozing off, head nods, slow blink rate, yawning while simulating a natural driving scenario. The videos are taken using an infrared (IR) camera that has a resolution of 640 X 480 and 30 frames per second. 

\paragraph{Classifier model preparation}
A deep learning model is utilized here to detect the driver's drowsy status. Individual frames in the video to carry out the prediction were used.

\begin{figure}[!t]
	\centering
	\includegraphics[width=.3\textwidth]{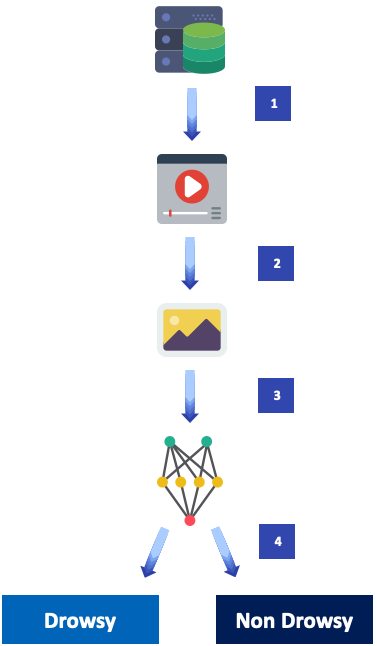}
	\caption{Pre-prossessing and training steps}\label{preprocessing steps}
\end{figure}

The creation of the model involves four steps as illustrated in figure \ref{preprocessing steps}.
\begin{itemize}
	\item Step 1– Selection of videos from the NTHU Database:\\
	The subjects selected for the model training needs to be of different ethnicities. Out of the selected participants, 18 subjects were part of the training dataset and 4 subjects were used in the evaluation dataset.
	\item Step 2 – Converting video into frames of images:\\
	The videos are converted into a series of images and each frame is represented as drowsy or non- drowsy.
	\item Step 3 – Training the algorithm:\\
	The series of images are the input to the algorithm, the algorithm is a CNN with three hidden layers as explained in Algorithm \ref{algorithm}. 
	During the training phase, the model learns about the drowsiness levels from the training dataset. The training will continue until the preferred level of accuracy is achieved for the model on the training data.
	
	\begin{algorithm}[!t]
		\caption{Real-Time Driver Drowsiness Detection Algorithm }
		\label{algorithm}
		\begin{algorithmic}[1]
			\item[\textbf{Input:} images of drivers and labels]
			\item[\textbf{Output:} Learned CNN model]
			\STATE Input (1X128X128) 
			\STATE Convolution   64(3X3)
			\STATE Leaky ReLU ($\alpha$ = 0.1)
			\STATE Max Pooling (2X2)
			\STATE Dropout (0.25)
			\STATE Convolution   128(3X3)
			\STATE Leaky ReLU ($\alpha$ = 0.1)
			\STATE Max Pooling (2 X2)
			\STATE Dropout (0.25)
			\STATE Convolution   128(3X3)
			\STATE Leaky ReLU ($\alpha$ = 0.1)
			\STATE Max Pooling (2X2)
			\STATE Dropout (0.25)
			\STATE Flatten 
			\STATE Dense (128 , activation=`linear')
			\STATE Leaky ReLU ($\alpha$ = 0.1)
			\STATE Dropout (0.5)
			\STATE Softmax Output
		\end{algorithmic}
	\end{algorithm}	
	
	\item Step 4 – Predictions and Saved Model:\\
	With the trained model, predictions can be made. This algorithm is saved onto the Android device to be used by the mobile application.
\end{itemize}

\subsubsection{Vehicle Data Collection System}
The vehicle data collection system is designed to collect trip data information i.e start time, end time, distance, and the minimum, maximum, and average speed, acceleration, distance, and GPS position every 15 seconds. It also collects information from the magnetometer in the phone to measure the rotational velocity along the Roll, Pitch and Yaw axes.

\begin{figure*}[!t]
	\centering
	\includegraphics[width=.8\textwidth]{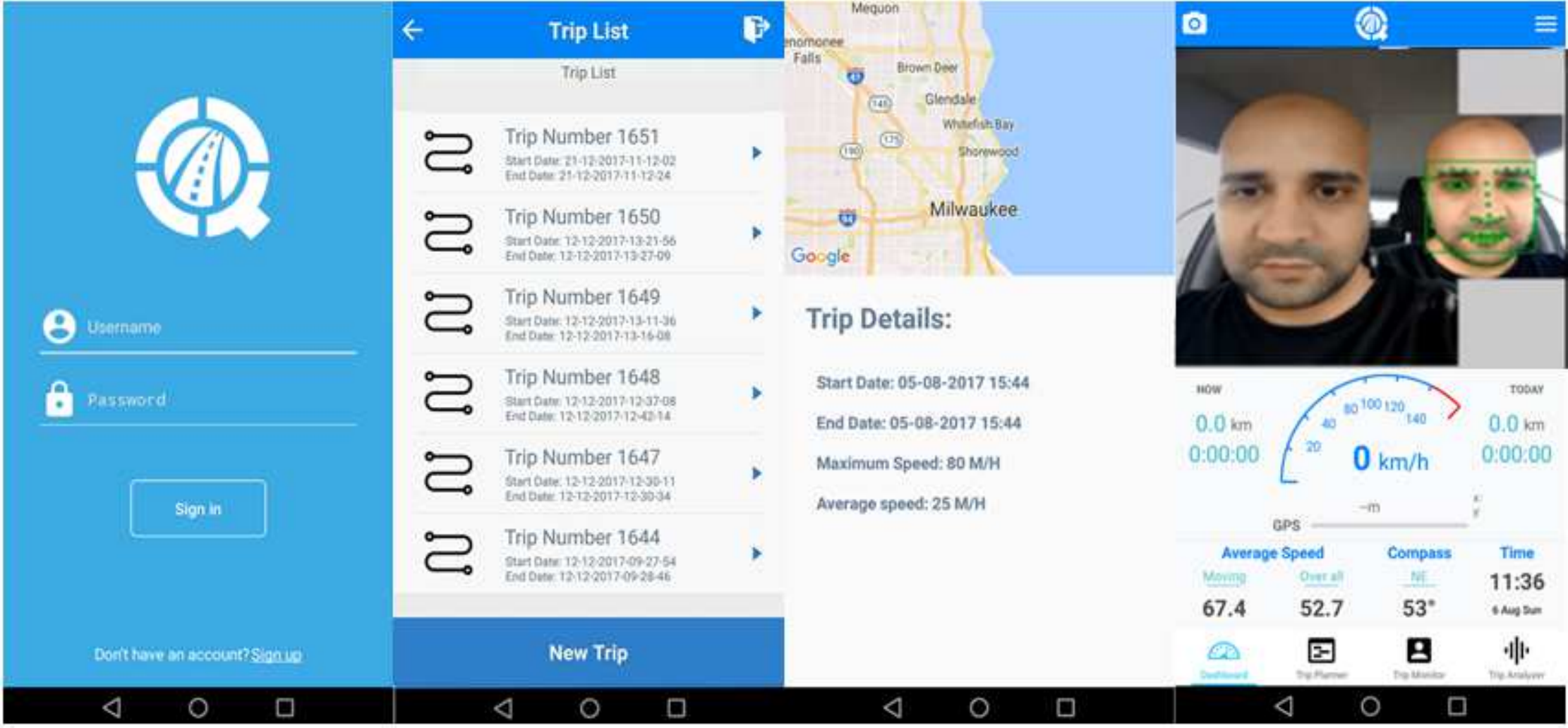}
	\caption{Screenshot of the four main pages in the Android application.}\label{application_pages}
\end{figure*}

As illustrated in figure.\ref{application_pages}, the Android application is composed of four main pages. The first page serves for logging in by using a username and password. Following the authentication, the user can start a new trip or access the information about the last trips.

If the user chooses a new trip, the application will start recording information about the driver's behavior, trip, and send the collected data via a web service to the database server. The current time and speed, the GPS position, duration of the trip, GPS precision, compass heading, minimum, average, and maximum speed, a rotational velocity along the Roll, Pitch and Yaw axes, and acceleration are also displayed. Video is captured using the front-facing camera and this is fed to the Machine Learning algorithm for drowsy driving detection.


\subsection{The network layer}
The network layer establishes the connection between the servers and transmits and processes the sensor data. The application can use either Wi-Fi or mobile internet (3G/3G+/4G) to send the data to the server. This process uses a hybrid system to gather and store data locally before transmitting them to the server. This technique is very effective when the internet connection is poor or unstable.

\subsection{The application layer}
The application layer delivers application-specific services to the end-user. It sends the obtained data to the web services for processing and analysis before showing them to the end-user. The web service is a component of the application layer. In addition to collecting data from the devices, it can also use sources such as the general traffic directory of the ministry to obtain information about crashes. This web service stores data in the database server and performs the analysis. User data can also be viewed by the end-user through this web interface. Windows Communication Foundation by Microsoft is used to implement the web service, based on the REST architecture and JSON format. As a result, the system can be easily set up to interact with any other embedded system. 

The Database server stores and queries the collected data. The database architecture used in the Database server consists of seven main tables. “Driver” table stores demographic information such as age and gender of the driver. The “Vehicle” section has information related to the driver's car. The road information is kept in the “Road” table. In the table “SegmentRoad,” the focus is on collecting data related to the road segments, which encompass the infrastructural information about the road. The table "Crashes" consists of the data retrieved from the stakeholders about historical crashes. This table is related to the table "Segment Road" so that it is possible to compute the risk related to different segments, as they are variable. Lastly, the "Events" table records the data that the mobile records every 15 seconds. The system is flexible in a way that you may keep adding variables to each table as the need arises in order to facilitate the computation of new proposed analytical models. Microsoft SQL Technologies are used to build the database server to ensure adequate performance, flexibility ,and scalability when dealing with extensive data in a cost-effective manner.


The Web application is the interface the researchers use to
interact and query the recorded data. The website displays demographic information about the driver. It also includes information about the vehicle such as the model and date of putting it into service.

\begin{figure*}
	\centering
	\includegraphics[width=.8\textwidth]{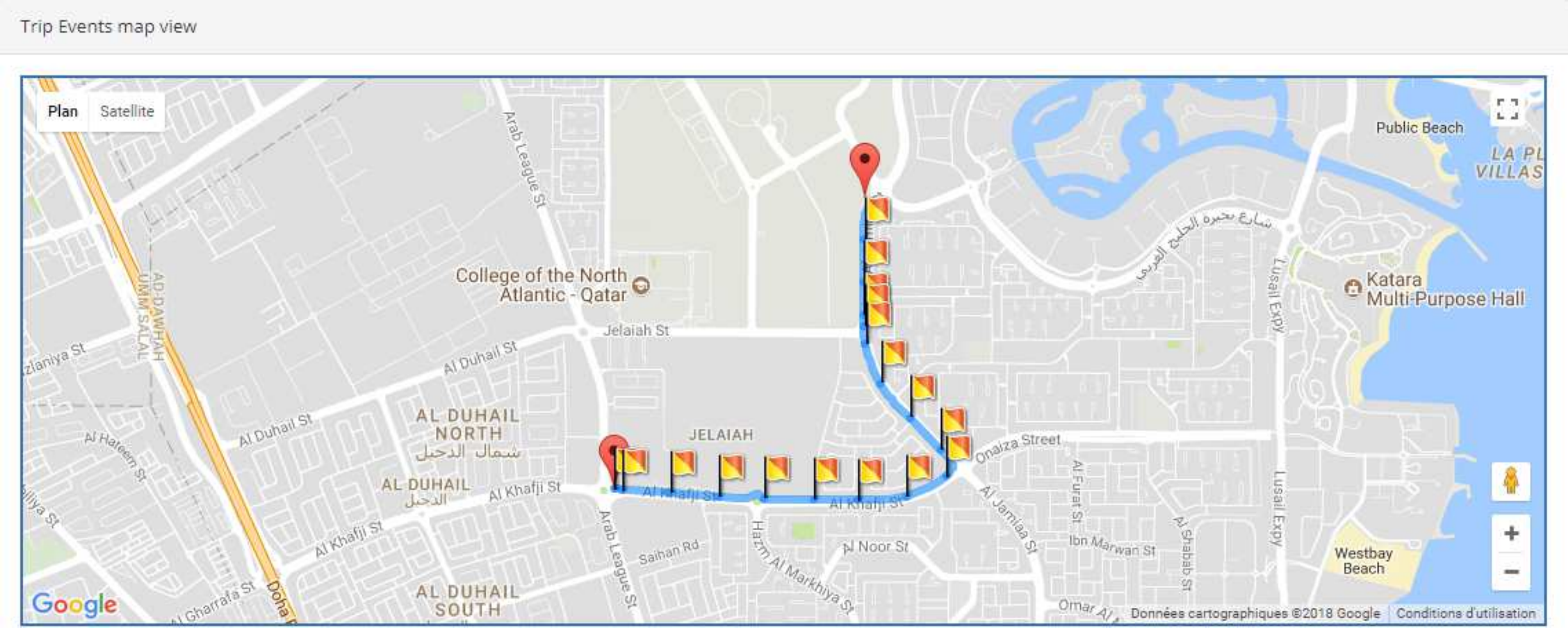}
	\caption{Screenshot of a real trip displaying the tracked trip and the position of individual events.}
	\label{fig:07}
\end{figure*}

\begin{figure*}
	\centering
	\includegraphics[width=.8\textwidth]{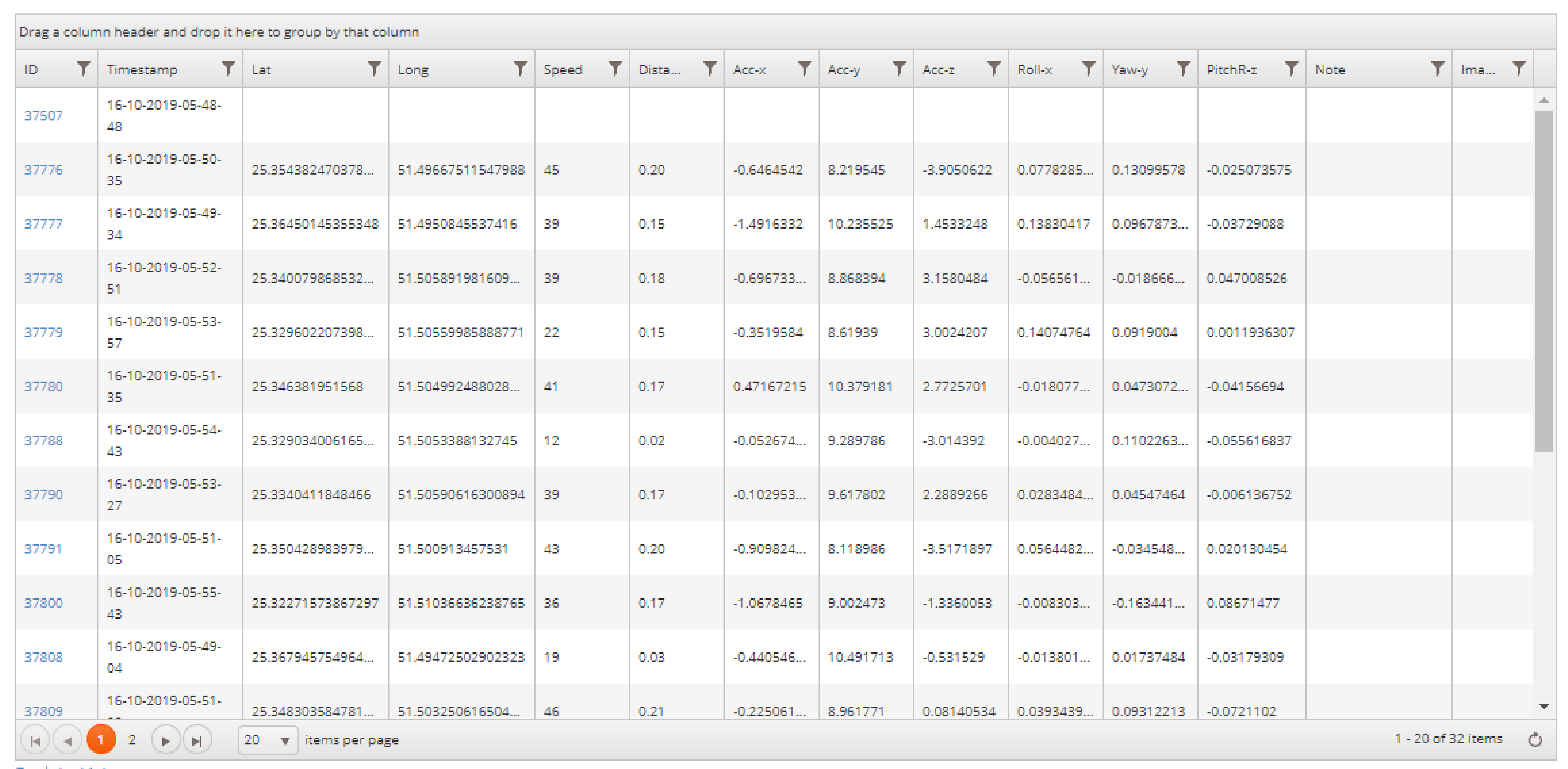}
	\caption{Screenshot of a real trip displaying the data recorded for every event.}\label{fig:08}
\end{figure*}

By using Google Maps API, the website displays the tracked trip and the position of individual events as illustrated figure \ref{fig:07}, as well as the details of all recorded events as shown in figure.\ref{fig:08}.

\section{Experimental Result}
For our driver drowsiness detection system, subjects were recorded in five simulated driving scenarios - with glasses, without glasses, with sunglasses, night without glasses and night with glasses. Every driving scenario has two different states - sleepy and non-sleepy. As a result, a total of 200 videos were used.

The number of videos used for each category in the training dataset was 36. However, the number of extracted images varies. In total, there were 50,991 images in the scenario recorded in "night with glasses", 52,372 performed in "night without glasses", 106,992 in "with glasses", 107, 990 in "with sunglasses" and 108,380 in "without glasses". In the evaluation set, each category has 4 videos with the following number of extracted images: 32,922 images in "the night with glasses" scenario, 29,781 "the night without glasses" scenario, 37,357 in "with glasses" scenario, 28,214 in "with sunglasses" scenario, and 45,005 in "without glasses" scenario. In total, 599,894 images were extracted all from the 200 videos.

The training on the neural network model was performed by the computer with the following specifications: Intel Core i7-7500U, 16 GB RAM, NVIDIA GeForce GTX 670MX. Table~\ref{tab:01} shows the results for the mentioned categories. According to the obtained results, the proposed solution attained on average for all scenarios an 83\% accuracy rate, which when compared to Park et al.\cite{Park2017} exceeds it and comes shy of 2 percentage points when compared to Guo et al. \cite{Guo2019} work.

\begin{table}[!t]
	\centering
	\caption{Accuracy per driving scenarios}
	\label{tab:01}
	\begin{tabular}{ll}
		\toprule
		Category              & Accuracy \\ \midrule
		With glasses          & 85.548   \\
		Night Without glasses & 83.20    \\
		Night With glasses    & 79.142   \\
		Without glasses       & 89.16 \\
		With sunglasses       & 78.115   \\
		All                   & 82.9  \\ \bottomrule
	\end{tabular}
\end{table}

On further inspection of the images that were detected and not detected, it was revealed that the most important facial feature for drowsiness classification in all situations were the eyes. As a consequence, it is not surprising that wearing sunglasses decreases the efficiency of the model, as the algorithm does not take into consideration the driver's eyes. Another crucial criterion for better performance is the brightness of the image. Indeed, when the image is brighter, the error rate decreases by 6\%.

\section{Conclusion}
In this paper, we presented a platform based on the Internet of Things to examine and analyse the traffic in the State of Qatar. The system focuses on implementing a naturalistic driver behavior approach and includes three primary layers. The perception layer represented in the Android application that includes the detector system of the behavior of the driver and the vehicle data collection system in order to collect data about the trip and the vehicle. Second, the network layer establishes the connection between the smartphone and the cloud. In the third layer, the web server obtains and analyzes data collected from the sensors, the database server stores the collected data and the website displays the data and analysis. The platform has been tested using two mobile phones and relevant variables were recorded. Results confirm a good performance of the drowsiness detection since the average accuracy is more than 82\% as shown in table~\ref{tab:01} for all scenarios.

\section{Future Work}

The current IoT-based system can be improved in several scenarios by porting many functionalities of ADAS into a portable modular Android device in the context of the Internet of Vehicles. This application can make use of the rear camera on the mobile device and include information from the environment too. For example, data from traffic signals, road signals can also be identified to provide feedback to the driver. The ADAS will be able to detect pedestrians and other vehicles and warn the driver of a potential collision.

\section*{Acknowledgement}
This publication was funded by the NPRP award [NPRP8-910-2-387] from Qatar National Research Fund (a member of Qatar Foundation). The statements made herein are solely the responsibility of the authors.

\bibliographystyle{IEEEtran}
\bibliography{main}

\begin{thebibliography}{10}
\providecommand{\url}[1]{#1}
\csname url@samestyle\endcsname
\providecommand{\newblock}{\relax}
\providecommand{\bibinfo}[2]{#2}
\providecommand{\BIBentrySTDinterwordspacing}{\spaceskip=0pt\relax}
\providecommand{\BIBentryALTinterwordstretchfactor}{4}
\providecommand{\BIBentryALTinterwordspacing}{\spaceskip=\fontdimen2\font plus
\BIBentryALTinterwordstretchfactor\fontdimen3\font minus
  \fontdimen4\font\relax}
\providecommand{\BIBforeignlanguage}[2]{{%
\expandafter\ifx\csname l@#1\endcsname\relax
\typeout{** WARNING: IEEEtran.bst: No hyphenation pattern has been}%
\typeout{** loaded for the language `#1'. Using the pattern for}%
\typeout{** the default language instead.}%
\else
\language=\csname l@#1\endcsname
\fi
#2}}
\providecommand{\BIBdecl}{\relax}
\BIBdecl

\bibitem{who2019}
\BIBentryALTinterwordspacing
``{Road traffic injuries}.'' [Online]. Available:
  \url{https://www.who.int/en/news-room/fact-sheets/detail/road-traffic-injuries}
\BIBentrySTDinterwordspacing

\bibitem{ministry2018}
\BIBentryALTinterwordspacing
``A press conference of ministry of interior about the traffic situation during
  2017 in qatar - ministry of interior postal.'' [Online]. Available:
  \url{t.ly/jz0bV}
\BIBentrySTDinterwordspacing

\bibitem{udrive}
\BIBentryALTinterwordspacing
``{What is NDS - UDRIVE, European Naturalistic Driving Study}.'' [Online].
  Available:
  \url{http://www.udrive.eu/index.php/about-udrive/what-is-naturalistic-driving}
\BIBentrySTDinterwordspacing

\bibitem{VanSchagen2012}
I.~van Schagen and F.~Sagberg, ``{The Potential Benefits of Naturalistic
  Driving for Road Safety Research: Theoretical and Empirical Considerations
  and Challenges for the Future},'' \emph{Procedia - Social and Behavioral
  Sciences}, vol.~48, pp. 692--701, 2012.

\bibitem{Dingus2015}
T.~Dingus, J.~Hankey, J.~Antin, S.~Lee, L.~Eichelberger, K.~E. Stulce,
  D.~McGraw, M.~Perez, and L.~Stowe, ``{Naturalistic Driving Study: Technical
  Coordination and Quality Control (No. SHRP 2 Report S2-S06-RW-1)},'' Virginia
  Tech Transportation Institute, Tech. Rep., 2015.

\bibitem{Ingrid2010}
S.~Ingrid, E.~Rob, and N.~Nicole, ``{Promoting real life observations for
  gaining understanding of road user behavior in Europe},'' \emph{Proceedings
  of the Road Safety on Four Continents Conference}, vol.~15, pp. 200--207,
  2010.

\bibitem{campbell2012shrp}
K.~L. Campbell, ``The shrp 2 naturalistic driving study: Addressing driver
  performance and behavior in traffic safety,'' \emph{Tr News}, no. 282, 2012.

\bibitem{Nahrstedt2016}
K.~Nahrstedt, H.~Li, P.~Nguyen, S.~Chang, and L.~Vu, ``{Internet of mobile
  things: Mobility-driven challenges, designs and implementations},'' in
  \emph{Proceedings - 2016 IEEE 1st International Conference on
  Internet-of-Things Design and Implementation, IoTDI 2016}.\hskip 1em plus
  0.5em minus 0.4em\relax Institute of Electrical and Electronics Engineers
  Inc., may 2016, pp. 25--36.

\bibitem{He2015}
\BIBentryALTinterwordspacing
K.~He, X.~Zhang, S.~Ren, and J.~Sun, ``{Deep residual learning for image
  recognition},'' Tech. Rep., 2016. [Online]. Available:
  \url{http://image-net.org/challenges/LSVRC/2015/}
\BIBentrySTDinterwordspacing

\bibitem{Long}
E.~Shelhamer, J.~Long, and T.~Darrell, ``{Fully Convolutional Networks for
  Semantic Segmentation},'' Tech. Rep.~4, 2017.

\bibitem{vijayan2019}
V.~Vijayan and E.~Sherly, ``{Real time detection system of driver drowsiness
  based on representation learning using deep neural networks},'' in
  \emph{Journal of Intelligent and Fuzzy Systems}, vol.~36, no.~3.\hskip 1em
  plus 0.5em minus 0.4em\relax IOS Press, 2019, pp. 1977--1985.

\bibitem{Dwivedi2014}
K.~Dwivedi, K.~Biswaranjan, and A.~Sethi, ``{Drowsy driver detection using
  representation learning},'' in \emph{Souvenir of the 2014 IEEE International
  Advance Computing Conference, IACC 2014}.\hskip 1em plus 0.5em minus
  0.4em\relax IEEE Computer Society, 2014, pp. 995--999.

\bibitem{liu2019}
W.~Liu, J.~Qian, Z.~Yao, X.~Jiao, and J.~Pan, ``{Convolutional two-stream
  network using multi-facial feature fusion for driver fatigue detection},''
  \emph{Future Internet}, vol.~11, no.~5, 2019.

\bibitem{Garcia-Garcia2018}
\BIBentryALTinterwordspacing
M.~Garc{\'{i}}a-Garc{\'{i}}a, A.~Caplier, M.~Rombaut, and M.~Rombaut, ``{Sleep
  Deprivation Detection for Real-Time Driver Monitoring using Deep Learning},''
  Tech. Rep., 2018. [Online]. Available:
  \url{https://hal.archives-ouvertes.fr/hal-01837080}
\BIBentrySTDinterwordspacing

\bibitem{Jabbar2018}
R.~Jabbar, K.~Al-Khalifa, M.~Kharbeche, W.~Alhajyaseen, M.~Jafari, and
  S.~Jiang, ``{Real-time Driver Drowsiness Detection for Android Application
  Using Deep Neural Networks Techniques},'' in \emph{Procedia Computer
  Science}, vol. 130.\hskip 1em plus 0.5em minus 0.4em\relax Elsevier B.V.,
  2018, pp. 400--407.

\bibitem{Weng2017}
C.~H. Weng, Y.~H. Lai, and S.~H. Lai, ``{Driver drowsiness detection via a
  hierarchical temporal deep belief network},'' in \emph{Lecture Notes in
  Computer Science (including subseries Lecture Notes in Artificial
  Intelligence and Lecture Notes in Bioinformatics)}, vol. 10118 LNCS.\hskip
  1em plus 0.5em minus 0.4em\relax Springer Verlag, 2017, pp. 117--133.

\bibitem{Park2017}
S.~Park, F.~Pan, S.~Kang, and C.~D. Yoo, ``{Driver drowsiness detection system
  based on feature representation learning using various deep networks},'' in
  \emph{Lecture Notes in Computer Science (including subseries Lecture Notes in
  Artificial Intelligence and Lecture Notes in Bioinformatics)}, vol. 10118
  LNCS.\hskip 1em plus 0.5em minus 0.4em\relax Springer Verlag, 2017, pp.
  154--164.

\bibitem{Guo2019}
J.~M. Guo and H.~Markoni, ``{Driver drowsiness detection using hybrid
  convolutional neural network and long short-term memory},'' \emph{Multimedia
  Tools and Applications}, vol.~78, no.~20, pp. 29\,059--29\,087, oct 2019.

\end{thebibliography}

\newpage{\pagestyle{empty}\cleardoublepage}
\end{document}